\documentclass[conference]{IEEEtran}
\IEEEoverridecommandlockouts
\usepackage{cite}
\usepackage{amsmath,amssymb,amsfonts}
\usepackage{algorithmic}
\usepackage{graphicx}
\usepackage{textcomp}
\usepackage{xcolor}
\usepackage{acronym}
\usepackage{subcaption}
\usepackage{threeparttable}

\def\BibTeX{{\rm B\kern-.05em{\sc i\kern-.025em b}\kern-.08em
    T\kern-.1667em\lower.7ex\hbox{E}\kern-.125emX}}
\begin{document}

\title{A 120dB Programmable-Range On-Chip Pulse Generator for Characterizing Ferroelectric Devices
\thanks{This work received funding from the European Union's H2020 research and innovation programme under the H2020 BeFerrosynaptic (871737), MeM-Scales (871371) projects.}
}
\author{\IEEEauthorblockN{
Shyam Narayanan\IEEEauthorrefmark{1},
Erika Covi\IEEEauthorrefmark{2}, 
Viktor Havel\IEEEauthorrefmark{2},  
Charlotte Frenkel \IEEEauthorrefmark{1}, 
Suzanne Lancaster \IEEEauthorrefmark{2}, \\
Quang Duong \IEEEauthorrefmark{2}, 
Stefan Slesazeck \IEEEauthorrefmark{2}
Thomas Mikolajick \IEEEauthorrefmark{2}, 
Melika Payvand \IEEEauthorrefmark{1}, 
Giacomo Indiveri \IEEEauthorrefmark{1}}\\
\IEEEauthorblockA{
\IEEEauthorrefmark{1}Institute of Neuroinformatics,
University of Zurich and ETH Zurich, Switzerland  \\
\IEEEauthorrefmark{2}NaMLab gGmbH, Dresden, Germany
\\
Email: \{shyam, melika, giacomo\}@ini.uzh.ch}
}
%\author{\IEEEauthorblockN{Shyam Narayanan}
%\IEEEauthorblockA{\textit{Institute of Neuroinformatics} \\
%\textit{University and ETH Zürich}\\
%Zürich, Switzerland \\
%shyam@ini.uzh.ch}
%\and
%\IEEEauthorblockN{x}
%\and
%\and
%\and
%\and

\newcommand{\MP}[1]{{\color{red}  Melika: #1}}
\newcommand{\GI}[1]{{\color{blue}  Giacomo: #1}}
\newcommand{\SN}[1]{{\color{orange}  Shyam: #1}}
\newcommand{\ST}[1]{{\color{forestgreen}  Stefan: #1}}
\newcommand{\EC}[1]{{\color{teal}  Erika: #1}}
\newcommand{\SL}[1]{{\color{violet}  Suzanne: #1}}

\acrodef{ADC}[ADC]{Analog to Digital Converter}
\acrodef{ADEXP}[AdExp-I\&F]{Adaptive-Exponential Integrate and Fire}
\acrodef{AER}[AER]{Address-Event Representation}
\acrodef{AEX}[AEX]{AER EXtension board}
\acrodef{AE}[AE]{Address-Event}
\acrodef{AFM}[AFM]{Atomic Force Microscope}
\acrodef{AGC}[AGC]{Automatic Gain Control}
\acrodef{AI}[AI]{Artificial Intelligence}
\acrodef{ALD}[ALD]{Atomic Layer Deposition}
\acrodef{AMDA}[AMDA]{AER Motherboard with D/A converters}
\acrodef{ANN}[ANN]{Artificial Neural Network}
\acrodef{API}[API]{Application Programming Interface}
\acrodef{ARM}[ARM]{Advanced RISC Machine}
\acrodef{ASIC}[ASIC]{Application Specific Integrated Circuit}
\acrodef{AdExp}[AdExp-IF]{Adaptive Exponential Integrate-and-Fire}
\acrodef{BCI}[BCI]{Brain-Computer-Interface}
\acrodef{BCM}[BCM]{Bienenstock-Cooper-Munro}
\acrodef{BD}[BD]{Bundled Data}
\acrodef{BEOL}[BEOL]{Back-end of Line}
\acrodef{BG}[BG]{Bias Generator}
\acrodef{BMI}[BMI]{Brain-Machince Interface}
\acrodef{BTB}[BTB]{band-to-band tunnelling}
\acrodef{BP}[BP]{Back-propagation}
\acrodef{BPTT}[BPTT]{Back-propagation Through Time}
\acrodef{CAD}[CAD]{Computer Aided Design}
\acrodef{CAM}[CAM]{Content Addressable Memory}
\acrodef{CAVIAR}[CAVIAR]{Convolution AER Vision Architecture for Real-Time}
\acrodef{CA}[CA]{Cortical Automaton}
\acrodef{CCN}[CCN]{Cooperative and Competitive Network}
\acrodef{CDAC}[CDAC]{Capacitive Digital to Analog Converter}
\acrodef{CDR}[CDR]{Clock-Data Recovery}
\acrodef{CFC}[CFC]{Current to Frequency Converter}
\acrodef{CHP}[CHP]{Communicating Hardware Processes}
\acrodef{CMIM}[CMIM]{Metal-insulator-metal Capacitor}
\acrodef{CML}[CML]{Current Mode Logic}
\acrodef{CMP}[CMP]{Chemical Mechanical Polishing}
\acrodef{CMOL}[CMOL]{Hybrid CMOS nanoelectronic circuits}
\acrodef{CMOS}[CMOS]{Complementary Metal-Oxide-Semiconductor}
\acrodef{CNN}[CCN]{Convolutional Neural Network}
\acrodef{COTS}[COTS]{Commercial Off-The-Shelf}
\acrodef{CPG}[CPG]{Central Pattern Generator}
\acrodef{CPLD}[CPLD]{Complex Programmable Logic Device}
\acrodef{CPU}[CPU]{Central Processing Unit}
\acrodef{CSM}[CSM]{Cortical State Machine}
\acrodef{CSP}[CSP]{Constraint Satisfaction Problem}
\acrodef{CV}[CV]{Coefficient of Variation}
\acrodef{DAC}[DAC]{Digital to Analog Converter}
\acrodef{DAS}[DAS]{Dynamic Auditory Sensor}
\acrodef{DAVIS}[DAVIS]{Dynamic and Active Pixel Vision Sensor}
\acrodef{DBN}[DBN]{Deep Belief Network}
\acrodef{DBS}[DBS]{Deep-Brain Stimulation}
\acrodef{DFA}[DFA]{Deterministic Finite Automaton}
\acrodef{DIBL}[DIBL]{drain-induced-barrier-lowering}
\acrodef{DI}[DI]{delay insensitive}
\acrodef{DMA}[DMA]{Direct Memory Access}
\acrodef{DNF}[DNF]{Dynamic Neural Field}
\acrodef{DNN}[DNN]{Deep Neural Network}
\acrodef{DoF}[DoF]{Degrees of Freedom}
\acrodef{DPE}[DPE]{Dynamic Parameter Estimation}
\acrodef{DPI}[DPI]{Differential Pair Integrator}
\acrodef{DRRZ}[DR-RZ]{Dual-Rail Return-to-Zero}
\acrodef{DRAM}[DRAM]{Dynamic Random Access Memory}
\acrodef{DR}[DR]{Dual Rail}
\acrodef{DSP}[DSP]{Digital Signal Processor}
\acrodef{DVS}[DVS]{Dynamic Vision Sensor}
\acrodef{DYNAP}[DYNAP]{Dynamic Neuromorphic Asynchronous Processor}
\acrodef{EBL}[EBL]{Electron Beam Lithography}
\acrodef{ECoG}[ECoG]{Electrocorticography}
\acrodef{EDVAC}[EDVAC]{Electronic Discrete Variable Automatic Computer}
\acrodef{EEG}[EEG]{electroencephalography}
\acrodef{EIN}[EIN]{Excitatory-Inhibitory Network}
\acrodef{EM}[EM]{Expectation Maximization}
\acrodef{EPSC}[EPSC]{Excitatory Post-Synaptic Current}
\acrodef{EPSP}[EPSP]{Excitatory Post-Synaptic Potential}
\acrodef{ET}[ET]{Eligibility Trace}
\acrodef{EZ}[EZ]{Epileptogenic Zone}
\acrodef{FDSOI}[FDSOI]{Fully-Depleted Silicon on Insulator}
\acrodef{FEOL}[FEOL]{Front-end of Line}
\acrodef{FET}[FET]{Field-Effect Transistor}
\acrodef{FFT}[FFT]{Fast Fourier Transform}
\acrodef{FI}[F-I]{Frequency-Current}
\acrodef{FPGA}[FPGA]{Field Programmable Gate Array}
\acrodef{FR}[FR]{Fast Ripple}
\acrodef{FSA}[FSA]{Finite State Automaton}
\acrodef{FSM}[FSM]{Finite State Machine}
\acrodef{GIDL}[GIDL]{gate-induced-drain-leakage}
\acrodef{GOPS}[GOPS]{Giga-Operations per Second}
\acrodef{GPU}[GPU]{Graphical Processing Unit}
\acrodef{GUI}[GUI]{Graphical User Interface}
\acrodef{HAL}[HAL]{Hardware Abstraction Layer}
\acrodef{HFO}[HFO]{High Frequency Oscillation}
\acrodef{HH}[H\&H]{Hodgkin \& Huxley}
\acrodef{HMM}[HMM]{Hidden Markov Model}
\acrodef{HRS}[HRS]{High-Resistive State}
\acrodef{HR}[HR]{Human Readable}
\acrodef{HSE}[HSE]{Handshaking Expansion}
\acrodef{HW}[HW]{Hardware}
\acrodef{IBCI}[IBCI]{Implantable BCI}
\acrodef{ICT}[ICT]{Information and Communication Technology}
\acrodef{IC}[IC]{Integrated Circuit}
\acrodef{ICL}[ICL]{Implantable Closed Loop}
\acrodef{IDAC}[IDAC]{Current Digital to Analog Converter}
\acrodef{IEEG}[iEEG]{intracranial electroencephalography}
\acrodef{IF2DWTA}[IF2DWTA]{Integrate \& Fire 2--Dimensional WTA}
\acrodef{IFSLWTA}[IFSLWTA]{Integrate \& Fire Stop Learning WTA}
\acrodef{IF}[I\&F]{Integrate-and-Fire}
\acrodef{IMU}[IMU]{Inertial Measurement Unit}
\acrodef{INCF}[INCF]{International Neuroinformatics Coordinating Facility}
\acrodef{INI}[INI]{Institute of Neuroinformatics}
\acrodef{INRC}[Intel NRC]{Intel Neuromorphic Research Community}
\acrodef{IO}[I/O]{Input/Output}
\acrodef{IoT}[IoT]{Internet of Things}
\acrodef{IPSC}[IPSC]{Inhibitory Post-Synaptic Current}
\acrodef{IPSP}[IPSP]{Inhibitory Post-Synaptic Potential}
\acrodef{IP}[IP]{Intellectual Property}
\acrodef{ISI}[ISI]{Inter-Spike Interval}
\acrodef{IoT}[IoT]{Internet of Things}
\acrodef{JFLAP}[JFLAP]{Java - Formal Languages and Automata Package}
\acrodef{LEDR}[LEDR]{Level-Encoded Dual-Rail}
\acrodef{LFP}[LFP]{Local Field Potential}
\acrodef{LFSR}[LFSR]{Linear Feedback Shift Register}
\acrodef{LIF}[LIF]{Leaky Integrate and Fire}
\acrodef{LLC}[LLC]{Low Leakage Cell}
\acrodef{LNA}[LNA]{Low-Noise Amplifier}
\acrodef{LPF}[LPF]{Low Pass Filter}
\acrodef{LRS}[LRS]{Low-Resistive State}
\acrodef{LSM}[LSM]{Liquid State Machine}
\acrodef{LTD}[LTD]{Long Term Depression}
\acrodef{LTI}[LTI]{Linear Time-Invariant}
\acrodef{LTP}[LTP]{Long Term Potentiation}
\acrodef{LTU}[LTU]{Linear Threshold Unit}
\acrodef{LUT}[LUT]{Look-Up Table}
\acrodef{LVDS}[LVDS]{Low Voltage Differential Signaling}
\acrodef{MD}[MD]{Medical Device}
\acrodef{MCMC}[MCMC]{Markov-Chain Monte Carlo}
\acrodef{MEMS}[MEMS]{Micro Electro Mechanical System}
\acrodef{MFR}[MFR]{Mean Firing Rate}
\acrodef{MIM}[MIM]{Metal Insulator Metal}
\acrodef{ML}[ML]{Machine Leanring}
\acrodef{MLP}[MLP]{Multilayer Perceptron}
\acrodef{MOSCAP}[MOSCAP]{Metal Oxide Semiconductor Capacitor}
\acrodef{MOSFET}[MOSFET]{Metal Oxide Semiconductor Field-Effect Transistor}
\acrodef{MOS}[MOS]{Metal Oxide Semiconductor}
\acrodef{MRI}[MRI]{Magnetic Resonance Imaging}
\acrodef{NDFSM}[NDFSM]{Non-deterministic Finite State Machine} 
\acrodef{ND}[ND]{Noise-Driven}
\acrodef{NEF}[NEF]{Neural Engineering Framework}
\acrodef{NHML}[NHML]{Neuromorphic Hardware Mark-up Language}
\acrodef{NIL}[NIL]{Nano-Imprint Lithography}
\acrodef{NLP}[NLP]{Natural Language Processing}
\acrodef{NMDA}[NMDA]{N-Methyl-D-Aspartate}
\acrodef{NME}[NE]{Neuromorphic Engineering}
\acrodef{NN}[NN]{Neural Network}
\acrodef{NRZ}[NRZ]{Non-Return-to-Zero}
\acrodef{NSM}[NSM]{Neural State Machine}
\acrodef{OR}[OR]{Operating Room}
\acrodef{OTA}[OTA]{Operational Transconductance Amplifier}
\acrodef{PCB}[PCB]{Printed Circuit Board}
\acrodef{PCHB}[PCHB]{Pre-Charge Half-Buffer}
\acrodef{PCM}[PCM]{Phase Change Memory}
\acrodef{PD}[PD]{Parkinson Disease}
\acrodef{PE}[PE]{Phase Encoding}
\acrodef{PFA}[PFA]{Probabilistic Finite Automaton}
\acrodef{PFC}[PFC]{prefrontal cortex}
\acrodef{PFM}[PFM]{Pulse Frequency Modulation}
\acrodef{PM}[PM]{Personalized Medicine}
\acrodef{PR}[PR]{Production Rule}
\acrodef{PSC}[PSC]{Post-Synaptic Current}
\acrodef{PSP}[PSP]{Post-Synaptic Potential}
\acrodef{PSTH}[PSTH]{Peri-Stimulus Time Histogram}
\acrodef{PVD}[PVD]{Physical Vapor Deposition }
\acrodef{QDI}[QDI]{Quasi Delay Insensitive}
\acrodef{RAM}[RAM]{Random Access Memory}
\acrodef{RDF}[RDF]{random dopant fluctuation}
\acrodef{RELU}[ReLu]{Rectified Linear Unit}
\acrodef{RLS}[RLS]{Recursive Least-Squares}
\acrodef{RMSE}[RMSE]{Root Mean Squared-Error}
\acrodef{RMS}[RMS]{Root Mean Squared}
\acrodef{RNN}[RNN]{Recurrent Neural Network}
\acrodef{ROLLS}[ROLLS]{Reconfigurable On-Line Learning Spiking}
\acrodef{RRAM}[R-RAM]{Resistive Random Access Memory}
\acrodef{R}[R]{Ripples}
\acrodef{SAC}[SAC]{Selective Attention Chip}
\acrodef{SAT}[SAT]{Boolean Satisfiability Problem}
\acrodef{SCX}[SCX]{Silicon CorteX}
\acrodef{SD}[SD]{Signal-Driven}
\acrodef{SDSP}[SDSP]{Spike Driven Synaptic Plasticity}
\acrodef{SEM}[SEM]{Spike-based Expectation Maximization}
\acrodef{SLAM}[SLAM]{Simultaneous Localization and Mapping}
\acrodef{SNN}[SNN]{Spiking Neural Network}
\acrodef{SNR}[SNR]{Signal to Noise Ratio}
\acrodef{SOC}[SOC]{System-On-Chip}
\acrodef{SOI}[SOI]{Silicon on Insulator}
\acrodef{SoA}[SoA]{state-of-the-art}
\acrodef{SP}[SP]{Separation Property}
\acrodef{SRAM}[SRAM]{Static Random Access Memory}
\acrodef{STDP}[STDP]{Spike-Timing Dependent Plasticity}
\acrodef{STD}[STD]{Short-Term Depression}
\acrodef{STEM}[STEM]{Science, Technology, Engineering and Mathematics}
\acrodef{STP}[STP]{Short-Term Plasticity}
\acrodef{STT-MRAM}[STT-MRAM]{Spin-Transfer Torque Magnetic Random Access Memory}
\acrodef{STT}[STT]{Spin-Transfer Torque}
\acrodef{SW}[SW]{Software}
\acrodef{TCAM}[TCAM]{Ternary Content-Addressable Memory}
\acrodef{TFT}[TFT]{Thin Film Transistor}
\acrodef{TPU}[TPU]{Tensor Processing Unit}
\acrodef{TRL}[TRL]{Technology Readiness Level}
\acrodef{USB}[USB]{Universal Serial Bus}
\acrodef{VHDL}[VHDL]{VHSIC Hardware Description Language}
\acrodef{VLSI}[VLSI]{Very Large Scale Integration}
\acrodef{VOR}[VOR]{Vestibulo-Ocular Reflex}
\acrodef{WCST}[WCST]{Wisconsin Card Sorting Test}
\acrodef{WTA}[WTA]{Winner-Take-All}
\acrodef{XML}[XML]{eXtensible Mark-up Language}
\acrodef{CTXCTL}[CTXCTL]{CortexControl}
\acrodef{divmod3}[DIVMOD3]{divisibility of a number by three}
\acrodef{hWTA}[hWTA]{hard Winner-Take-All}
\acrodef{sWTA}[sWTA]{soft Winner-Take-All}
\acrodef{APMOM}[APMOM]{Alternate Polarity Metal On Metal}
\maketitle

\begin{abstract}
Novel non-volatile memory devices based on ferroelectric thin films represent a promising emerging technology that is ideally suited for neuromorphic applications. The physical switching mechanism in such films is the nucleation and growth of ferroelectric domains. Since this has a strong dependence on both pulse width and voltage amplitude, it is important to use precise pulsing schemes for a thorough characterization of their behavior. In this work, we present an on-chip 120\,dB programmable range pulse generator, that can generate pulse widths ranging from 10\,ns to 10\,ms $\mathbf{\pm}$2.5\% which eliminates the RLC bottleneck in the device characterisation setup. We describe the pulse generator design and show how the pulse width can be tuned with high accuracy, using Digital to Analog converters. Finally, we present experimental results measured from the circuit, fabricated using a standard 180\,nm CMOS technology.
\end{abstract}


\section{Introduction}
Thanks to their ultra-low power operation, \ac{CMOS}-compatibility, and non-volatility, ferroelectric hafnium oxide-based devices such as ferroelectric capacitors (FeCAP)~\cite{Sony}, ferroelectric field effect transistors (FeFET)~\cite{GF}, and ferroelectric tunneling junctions (FTJ)~\cite{Benjamin2019} are widely explored for digital storage conventional memory array architectures~\cite{Sony,GF} as well as in novel bio-inspired architectures~\cite{Erika-ISCAS}. 
Besides multi-bit storage in digital memories, the analog switching properties of these nano-scale devices are important for the realization of synaptic weight elements~\cite{HalidSynapse,Mattia} or artificial neurons~\cite{HalidNeuron} for adoption in neuromorphic applications~\cite{Indiveri_Liu15}.
%Multi-bit resolution is particularly useful in neural networks during the training phase, since it increases the capacity of the network to learn more complex patterns~\cite{Yigitpaper}. 
In order to allow proper device optimization and defining the constraints for circuit design, a profound understanding of the switching kinetics in such nano-scale devices is needed~\cite{kinetics}. 
%To this end, complex characterization methods that use voltage pulses are an extremely valuable tool.

\begin{figure}
  \begin{subfigure}{0.5\textwidth}
    \centering
    \includegraphics[width=0.7\textwidth]{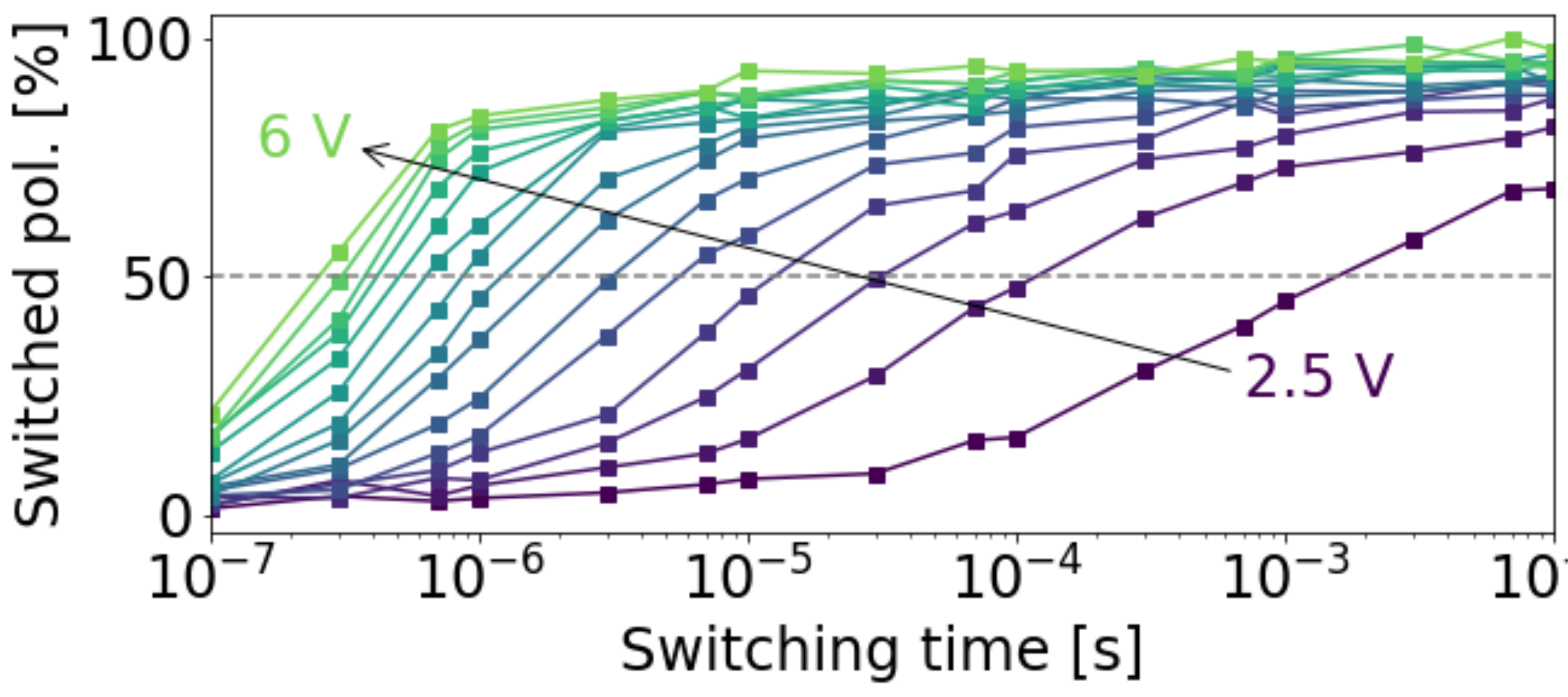}
    \subcaption{}
    \label{fig:switching_on}
  \end{subfigure}
  \begin{subfigure}{0.5\textwidth}
    \centering
    \includegraphics[width=0.71\textwidth]{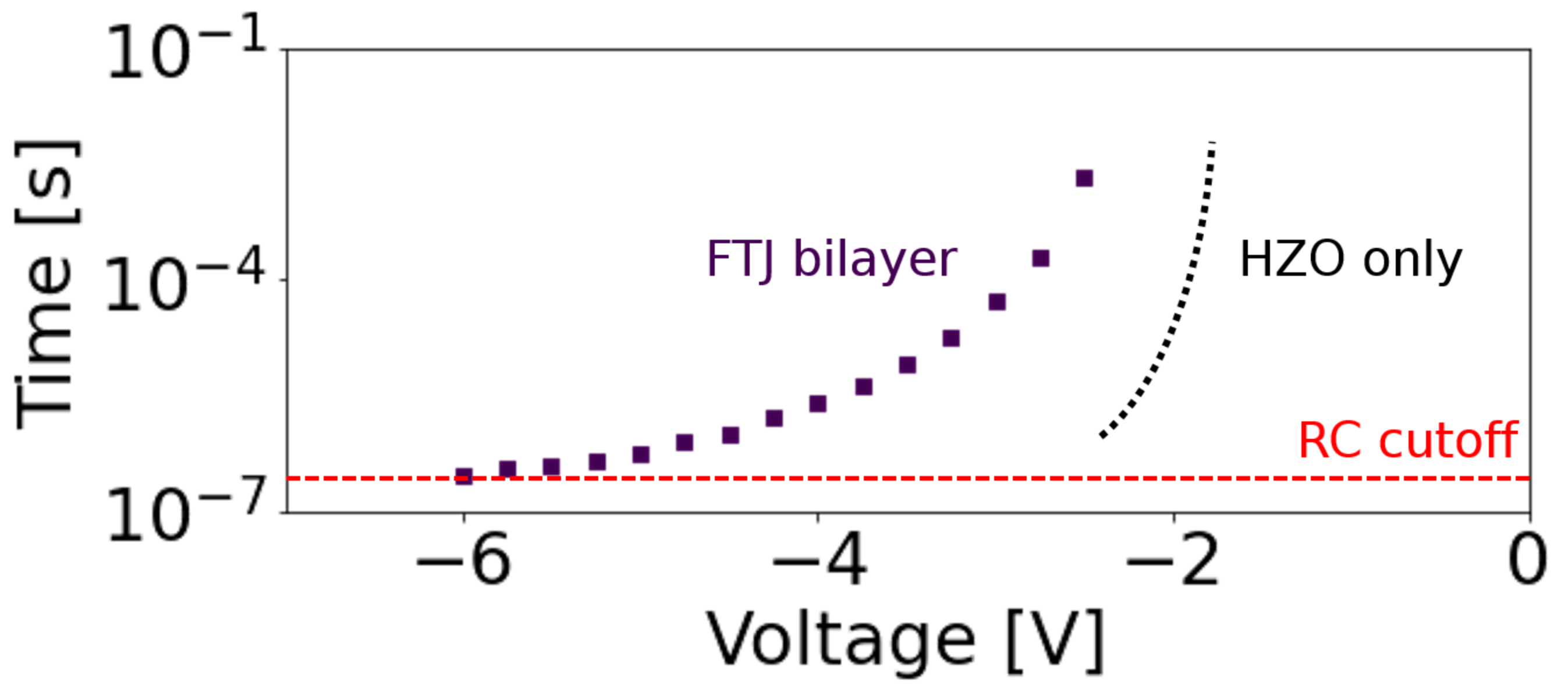}
    \subcaption{}
    \label{fig:switching_fitting}
  \end{subfigure}
  \caption{Switching transitions: (\subref{fig:switching_on})  switched polarization as a function of pulse width for different pulse amplitudes, for a bilayer FTJ device. (\subref{fig:switching_fitting}) Time needed to switch 50\% of polarization for each voltage amplitude, expected switching curve for HZO only (black dotted line from~\cite{materano2020polarization}) and asymptotic RC-cutoff (red dotted line)}
  \label{fig:switching}
  \vspace{-0.5cm} 
\end{figure}
Figure~\ref{fig:switching_on} depicts the results of such a switching kinetics measurement for a bilayer FTJ device featuring a 10\,nm ferroelectric hafnium-zirconium-oxide (HZO) and a 2\,nm alumnium oxide (Al$\mathrm{_2}$O$\mathrm{_3}$) layer sandwiched between two titanium nitride (TiN) electrodes~\cite{Benjamin2019}. The diagram depicts the portion of switched polarization upon application of different switching pulses. Extracting the respective switching voltage and time at $50\%$ switched polarization results in the switching kinetics curve depicted in Fig.~\ref{fig:switching_fitting}. Extrapolation of this data yields both the data retention time of the respective device~\cite{sw-ret} as well as the required pulse voltages for ns-switching times. The switching kinetics of a given device will be material- and size-dependent~\cite{kinetics}, and as such, the characterization should be performed on individual devices. As an example, when looking at the voltage-time dependence in Fig.~\ref{fig:switching_fitting} and comparing to the data for an HZO-only sample (extracted from ~\cite{materano2020polarization}), there is a shift to larger voltages due to the voltage drop over the Al$\mathrm{_2}$O$\mathrm{_3}$ layer, and an asymptotic saturation of switching time between 100\,ns and 1\,$\mathrm{\mu}$s. This saturation is related to a RC-delay within the FTJ samples and measurement set-up that cannot be easily avoided. Similar issues occur for FeFET and FeCAP single devices, even though switching times down to just 14\,ns to 20\,ns have been demonstrated in integrated memory arrays~\cite{Sony,GF}. Obviously, the RC-delay has a strong impact on the extrapolation of switching kinetics, especially at shorter time scales, and consequently a pulse generator is required which is able to perform such a thorough characterization by applying pulses with varying amplitudes and timewidths directly on-chip. Moreover, for the characterization of multi-bit and analogue storage a very precise control of the switching pulses is mandatory~\cite{covi2014TSM}, especially for the steep switching gradients at shorter pulse times (see Fig.~\ref{fig:switching_on}). In order to tackle this issue, we developed an optimized on-chip pulse-generator circuit, enabling the application of well-controlled voltage pulses with timings that span over six orders of magnitude.
% that can be used for the characterization of Back End Of Line (BEOL) integrated ferroelectric devices.
%The pulse-generator allows to apply well-controlled voltage pulses covering a wide range of voltages and timings that span over six orders of magnitude.

\section{System Architecture}
\label{sec:arch}

\begin{figure}
  \centering
  \includegraphics[width=0.45\textwidth]{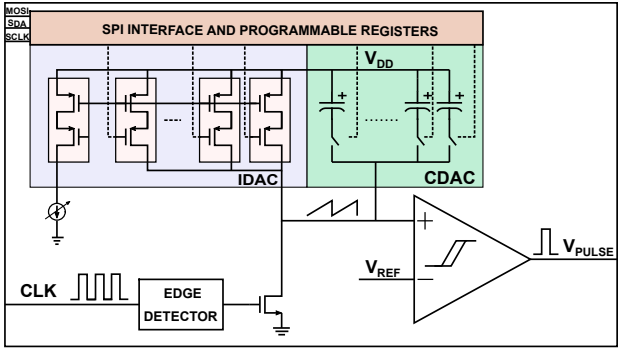}
  \caption{Architecture of the Pulse Generator}
  \label{fig:pg}
\end{figure}

Figure~\ref{fig:pg} shows the architecture of the wide-range programmable on-chip pulse generator. On the arrival of the input pulses (shown as $CLK$), the programmable capacitor $C_{DAC}$ is abruptly charged to $V_{DD}$, and once the pulse is removed, it slowly discharges through the $I_{DAC}$ current. This results in a ramp voltage on node $V_{RAMP}$ which when compared to a reference voltage ($V_{REF}$), generates a pulse at the output ($V_{PULSE}$). 
The slope of the ramp voltage is determined by the magnitude of the current $I_{DAC}$ and the value of $C_{DAC}$. Therefore, the width of the pulse at the output of the comparator can be determined by the following equation:

\begin{equation}
    I_{DAC} = C_{DAC} \cdot \frac{\Delta V_{REF}}{\Delta t} 
    \label{eq:1}
\end{equation}

To have a precise control on the output pulse width ($\Delta t$), the circuit should be independent of the input pulse width, and employ a high-precision capacitor $C_{DAC}$, high-precision discharge current $I_{DAC}$, and a low-offset comparator. In this section we describe each of these blocks.

\subsection{Edge detector}
To decouple the $V_{PULSE}$ pulse-width from that of the $CLK$ pulses, we implemented an Edge Detector block which detects the rising edge of the $CLK$ signal and sends a fixed-width pulse to the transistor $M_{PD}$, independent of the $CLK$ pulse width~\cite{weste2011cmos}. 
%The pulse width of the edge detector block depends on the delay on path XXX which is independent of the input. 

\subsection{Wide-Range \ac{IDAC}}
The wide-range IDAC block of Fig.~\ref{fig:pg} represents a 7-bit current output binary weighted \ac{DAC}. It consists of identical modular unit cells, combined in a binary weighted fashion to work as p-type current sources. The unit cells are sized appropriately to provide a precise value of current. Both p-type transistors where one being the current source and the other being the gating switch were sized to be of the same width but different lengths. The primary considerations for the sizing were the output resistance, matching, dynamic range and leakage depending on the function of each device.
The \ac{DAC} is biased by an external precise reference current which varies in three steps to cover the large operating range of the circuit as described in table \ref{tab:range}. 

One important consideration for the design of the current DAC is to ensure the current sources are accurate at large output currents, and that the current source transistors are not out of the saturation. The output impedance of the $I_{DAC}$ should be high to produce reliable output current with good accuracy. We decided to limit the systematic error $<1\%$ to help the system deliver a highly accurate output current. This is especially important, because as as shown in Figure~\ref{fig:pg}, the $I_{DAC}$  output is injected to the $V_{RAMP}$ node whose voltage is changing over time, and the $I_{DAC}$ current should not be susceptible to this change through the drain-source voltage across the current sources of the $I_{DAC}$. %Figure \ref{fig:compliance} describes this in detail. 
%The plot shows the output current at the highest $$I_{DAC}$$ value (binary code 1111111, equivalent to 127) among all the different operating $I_{DAC}$ currents described in table \ref{tab:range}. 
We verified that the systematic error in the  $I_{DAC}$  current is under 0.3\% of the desired value, if the operating headroom is at least 400-450\,mV.
%\begin{figure}[htbp]
%\centerline{\includegraphics[scale=0.43]{headroom.jpg}}
%\caption{$I_{DAC}$ Current Compliance}
%\label{fig:compliance}
%\end{figure}
\begin{table}
\center
\caption{$I_{DAC}$ Programmable Range}
\begin{tabular}{|c|c|l|}
\hline
\textbf{Pulse Range} & $\mathbf{I_{BIAS}}$ & \multicolumn{1}{c|}{$\mathbf{I_{DAC}}$} \\ \hline
7.8\,ns--1\,$\mu$s    & 160\,$\mu$A          & 10\,$\mu$A--1.28\,mA                                     \\ \hline
780\,ns--100\,$\mu$s  & 1.60\,$\mu$A         & 100\,nA--12.8\,$\mu$A                                    \\ \hline
78\,$\mu$s--10\,ms      & 16\,nA           & 1\,nA--128\,nA                                         \\ \hline
\end{tabular}
\label{tab:range}
\end{table}
We simulated the functionality of the circuit with a wide range of currents, and verified with experimental measurements the correct functionality of the circuit down to the lowest current setting of 1\,nA (see Section IV).
Typically, the circuit's systematic error is higher at lower currents, when the p-type transistors operate in weak inversion. However, narrow pulses are needed to characterize the memristive devices. Therefore, in this application scenario to generate narrow pulses, the required $I_{DAC}$ current is in the order of magnitude where the IDAC operates in strong inversion saturation.

\begin{figure}
  \centering
  \includegraphics[width=0.4\textwidth]{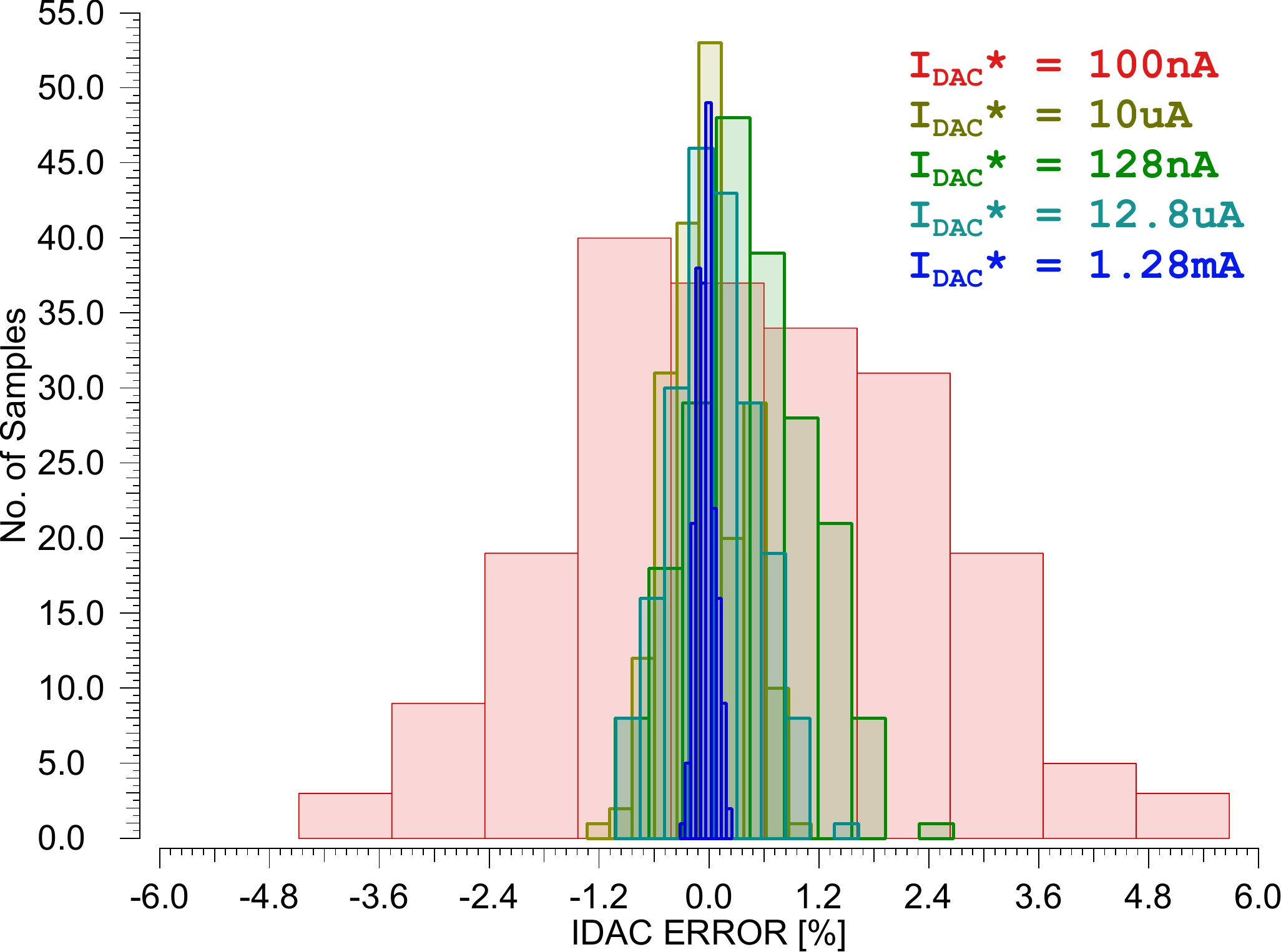}
  \caption{MonteCarlo Analysis of $I_{DAC}$ Error }
  \label{fig:idacmc}
\end{figure}

We performed a MonteCarlo simulation to quantify the expected accuracy and quality of this design, as we could not measure through a large set of silicon samples. In this analysis, we quantified the error for five different current settings ranging from the minimum to maximum of the 7-bit IDAC code across different range of operation as described in Table~\ref{tab:range}.
The deviation from the target value of the programmed current was calculated as:
\begin{equation}
    I_{DAC, Error} [\%] = \Big( \frac{I_{DAC,  Sim}}{I_{DAC^{*}}} -1 \Big)\cdot 100,
    \label{eq:2}
\end{equation}
where $I_{DAC,Sim}$ is the actual simulated $I_{DAC}$ output current and $I_{DAC^{*}}$ is the desired output current programmed with the 7-bit code. 
The results are shown in Fig.~\ref{fig:idacmc}. For this simulation, we limited the lower current setting of $I_{DAC^{*}}$ to 100\,nA as the MonteCarlo simulation in this technology yields a very large systematic offset (probably arising from inaccurate sub-threshold mismatch models). It can be observed from Fig.~\ref{fig:idacmc} that the lower current setting of 100\,nA shows the largest error, and the highest current setting of 1.28\,mA (required for the narrowest pulse) shows the least error, indicative of its high accuracy. 
Overall, the combined $3\sigma$ error of $I_{DAC}$ is within $\pm$ 6\% for all the current ranges tested, without any calibration process. Extra tuning for higher accuracy is possible by adjusting the capacitance value (trimming).

\subsection{$C_{DAC}$}
The primary capacitor in this system is an 8-bit binary weighted capacitor DAC ($C_{DAC}$). The $C_{DAC}$ is constructed out of identical unit DAC cells, carefully laid out to provide a good level of matching among all the elements in the array. The resolution of the $C_{DAC}$ is 62.5\,fF and the total capacitance we chose for our application is 10\,pF. This enables us to adjust the DAC value and correct/trim out any non-idealities in the design with a precision of $\pm 0.625\%$. As the desired value of 10\,pF (maximum $C_{DAC}$ value = 16pF) is not exactly at the centre of the DAC code range, we have non-uniform correction range centered around the desired value. However, one could potentially choose the exact center value of this DAC to be the capacitance required for pulse generation by either adjusting the $I_{DAC}$ current or the reference voltage $V_{REF}$ as described in Eq. ~\ref{eq:1}. 

\subsection{Hysteretic Comparator}
\begin{figure}
  \centering
  \includegraphics[width=0.3\textwidth]{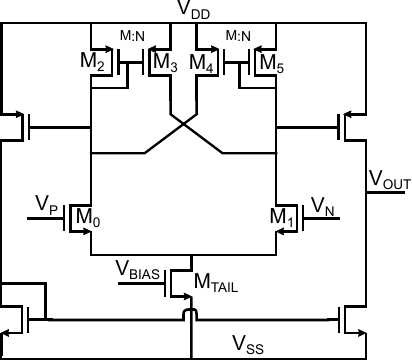}
  \caption{Differential to single- ended hysteretic comparator}
  \label{fig:comp}
\end{figure}

For this application, the primary considerations for a comparator was to have hysteresis to avoid false triggering due to fluctuations on the node the comparator monitors. In addition to this the other main considerations were to have low-offset and high-speed. For the comparator design, we chose a well-known architecture of a hysteretic comparator~\cite{AllstotComparator,allen2011cmos} as shown in Fig. \ref{fig:comp}. Positive feedback is introduced in this circuit to have hysteresis.

%In this circuit, there are two paths of feedback. The first is negative current-series feedback through the common-source node of transistors M0 and M1. In addition, there is a positive voltage-shunt feedback path through the gate–drain connections of transistors M3 and M4. To have hysteresis the positive feedback should be larger than the negative one such that the overall feedback in the system is positive.

\section{Implementation and system level simulation}
\begin{figure}
  \centering
  \includegraphics[width=0.35\textwidth]{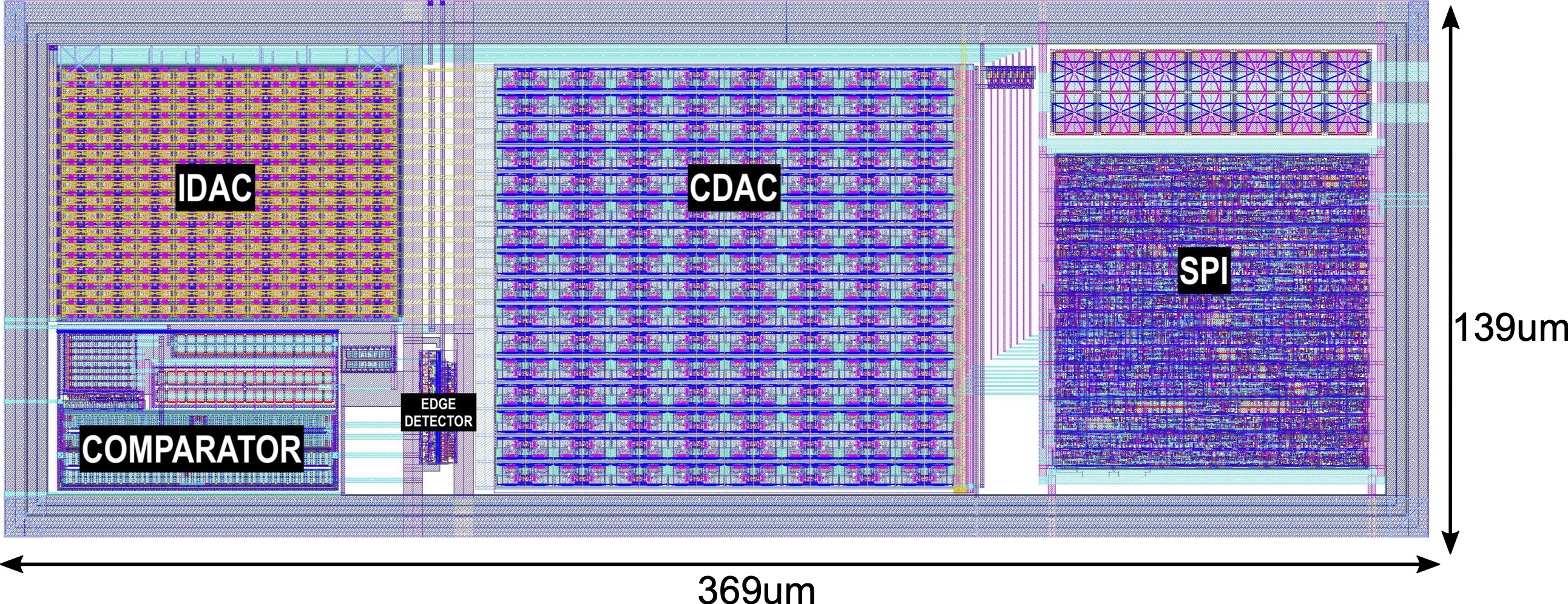}
  \caption{Pulse generator layout}
  \label{fig:pglay}
 
\end{figure}

\begin{figure}
  \centering
  \includegraphics[width=0.5\textwidth]{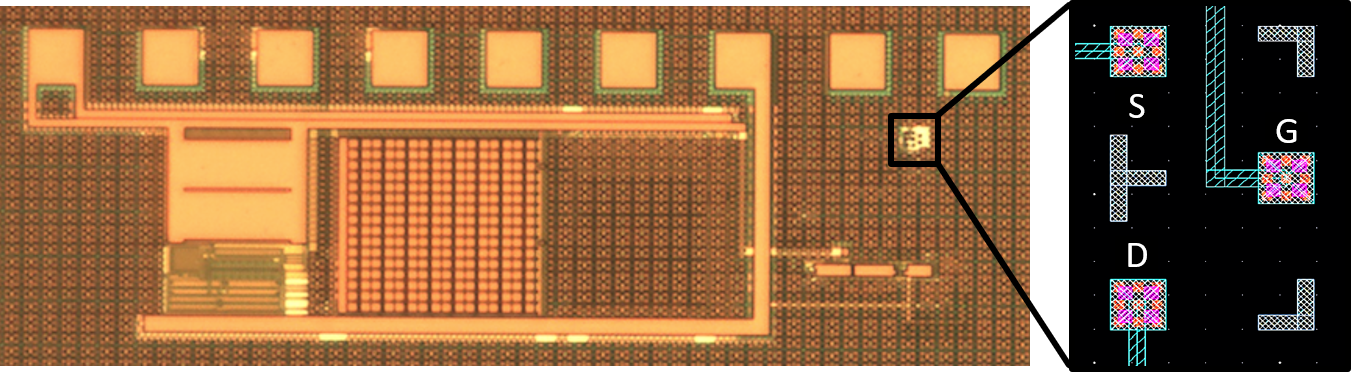}
\caption{Left: pulse generator die photograph. Right: Zoom of the contacts for ferroelectric device BEOL integration.}
\label{fig:pgphoto}
\end{figure}

The pulse generator was designed and fabricated in a standard 180\,nm technology node. The final design excluding the probe pads occupy an area of 369\,$\mu$m$\times$139\,$\mu$m as shown in Fig.~\ref{fig:pglay}. The chip microphotograph is shown in Fig.~\ref{fig:pgphoto}.

The pulse generator can be programmed using internal Serial Peripheral Interface (SPI) registers. Both $I_{DAC}$ and $C_{DAC}$ are internally wired to their respective registers which are tightly integrated with rest of the analog circuitry as shown in Fig.~\ref{fig:pglay}.
The I/Os are wired up to metal probe pads. They are powered and clocked externally through the wafer probing equipment. The FTJs will be fabricated on the same die with Back End Of Line (BEOL) post-processing integration steps. Therefore, the output of the pulse generator is directly connected to the FTJs with on chip low parasitic interconnects and switches.

Table~\ref{tab:perf} compares the performance of this work against the  implementations designed for similar application. This design can reliably cover a programmable range at least 4 orders of magnitude more than \cite{LeePG}, the digital implementation on the same technology node. Although compared to \cite{ErfanijaziPG}, the area benefit of our work is explainable by the smaller technology node and the programming range is 1 order of magnitude less, the accuracy of the pulses in our design across the full programmable range is $\mathbf{\pm}$2.5\%  which is not reported in \cite{LeePG} and \cite{ErfanijaziPG}. 
\begin{table} [h]
    \caption{Performance comparison}
    \renewcommand\arraystretch{1.5}
    \centering
\begin{threeparttable}    
\begin{tabular}{|c|c|c|c|}
    \hline
    \rule[-2mm]{0mm}{0.4cm}
    \textbf{Parameters}& 
    \textbf{This work}&
    \textbf{\cite{ErfanijaziPG}}&
    \textbf{\cite{LeePG}}\\
    \hline
    %\rule[-2mm]{0mm}{0.4cm}
    Technology [nm] & 180 & 350 & 180\\
    Implementation & Analog & Mixed & Digital\\
    Silicon Area [mm$^2$] & 0.051 & 0.203 & 0.015 \\
    Programmable Range [dB] & 120 & 150 & 42\\
    Output Pulse Accuracy [\%] & $\mathbf{\pm}$2.5 & N.R \tnote{*} & N.R \tnote{*}\\
    \hline
\end{tabular}
\footnotesize
\begin{tablenotes}
\item[*] Not Reported
\end{tablenotes}
\end{threeparttable}
%\vspace{-0.3cm}
\label{tab:perf}
\end{table}
\section{Hardware setup and electrical characterization}

%\EC{Responsible for the Section}
%\ST{perhaps add to the die photograph a small image form the layout showing the three contacts sticking out to BEOL-integrate the different ferroelectric devices}
%\EC{Done. I prepared two versions, I leave here the most compact. }

\begin{figure}
\centering
\includegraphics[width=0.4\textwidth]{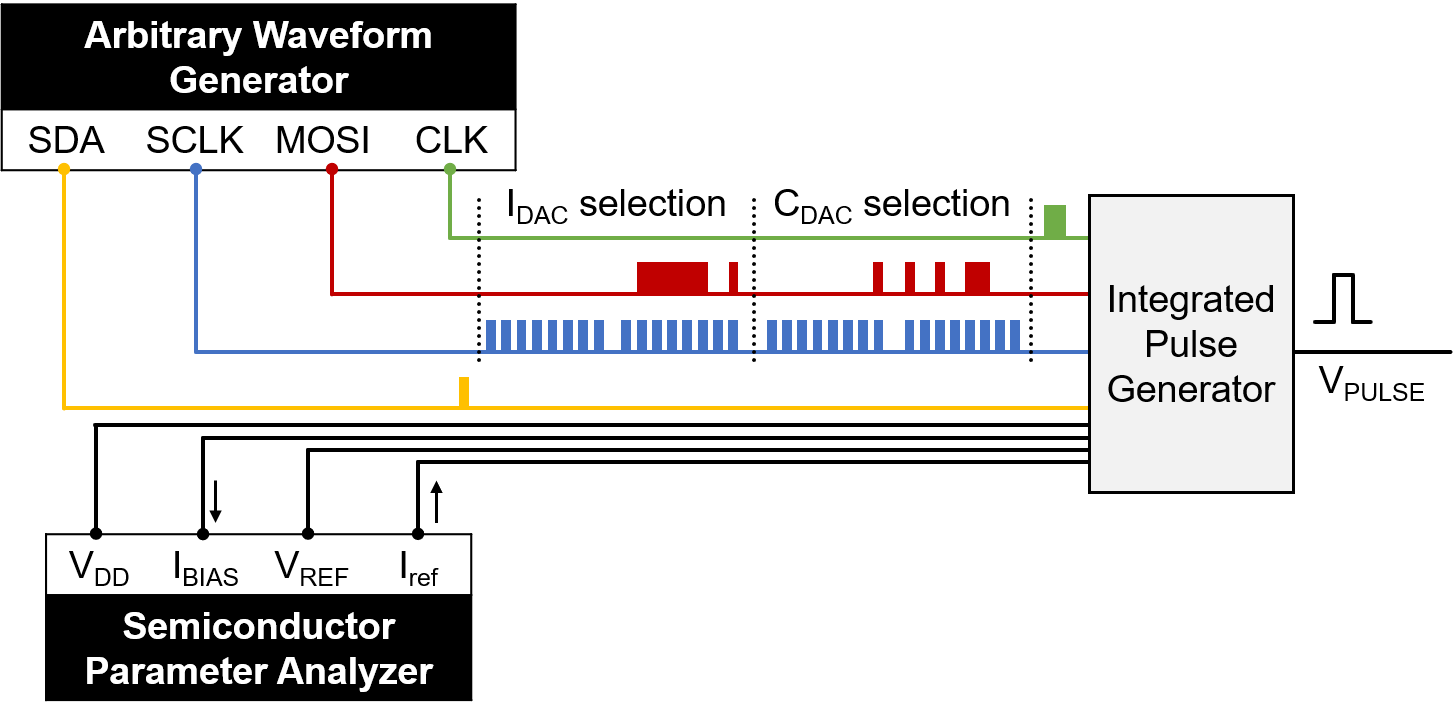}
\caption{Schematic of the electrical setup that provides the biases and signals to operate the pulse generator.}
\label{fig:setup}
\vspace{-0.3cm}
\end{figure}

To measure data from the fabricated pulse generator circuit we used a probe-station and applied constant voltages and currents using a Semiconductor Parameter Analyzer (see Fig.~\ref{fig:setup}). The SPI and CLK signals of Fig.~\ref{fig:setup} were generated using an Arbitrary Waveform Generator.

\begin{figure}[h]
  \begin{subfigure}{0.5\textwidth}
    \centering
    \includegraphics[width=0.85\textwidth]{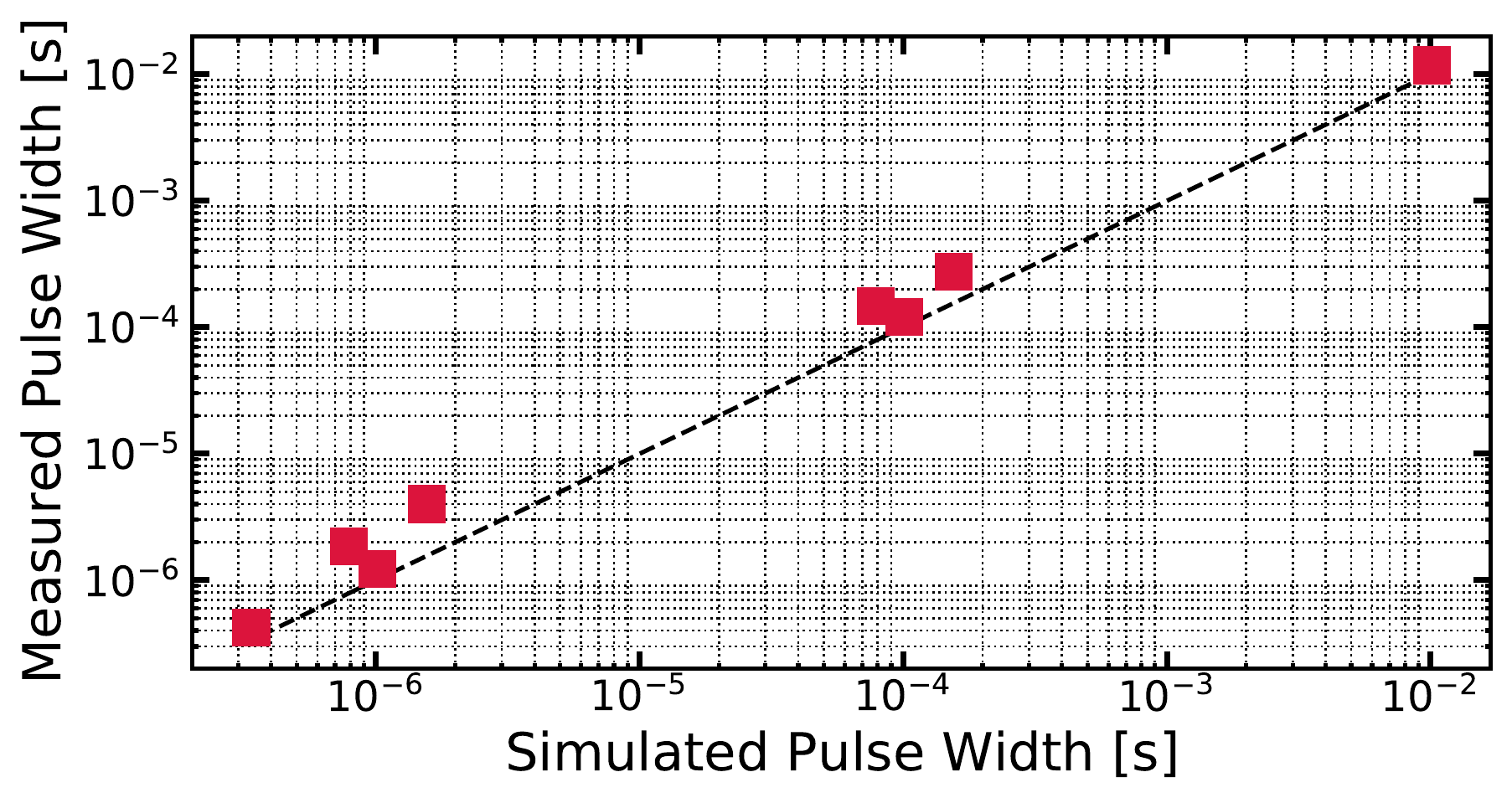}
    \subcaption{}
    \label{fig:measdataA}
  \end{subfigure}
  \begin{subfigure}{0.5\textwidth}
    \centering
    \includegraphics[width=0.85\textwidth]{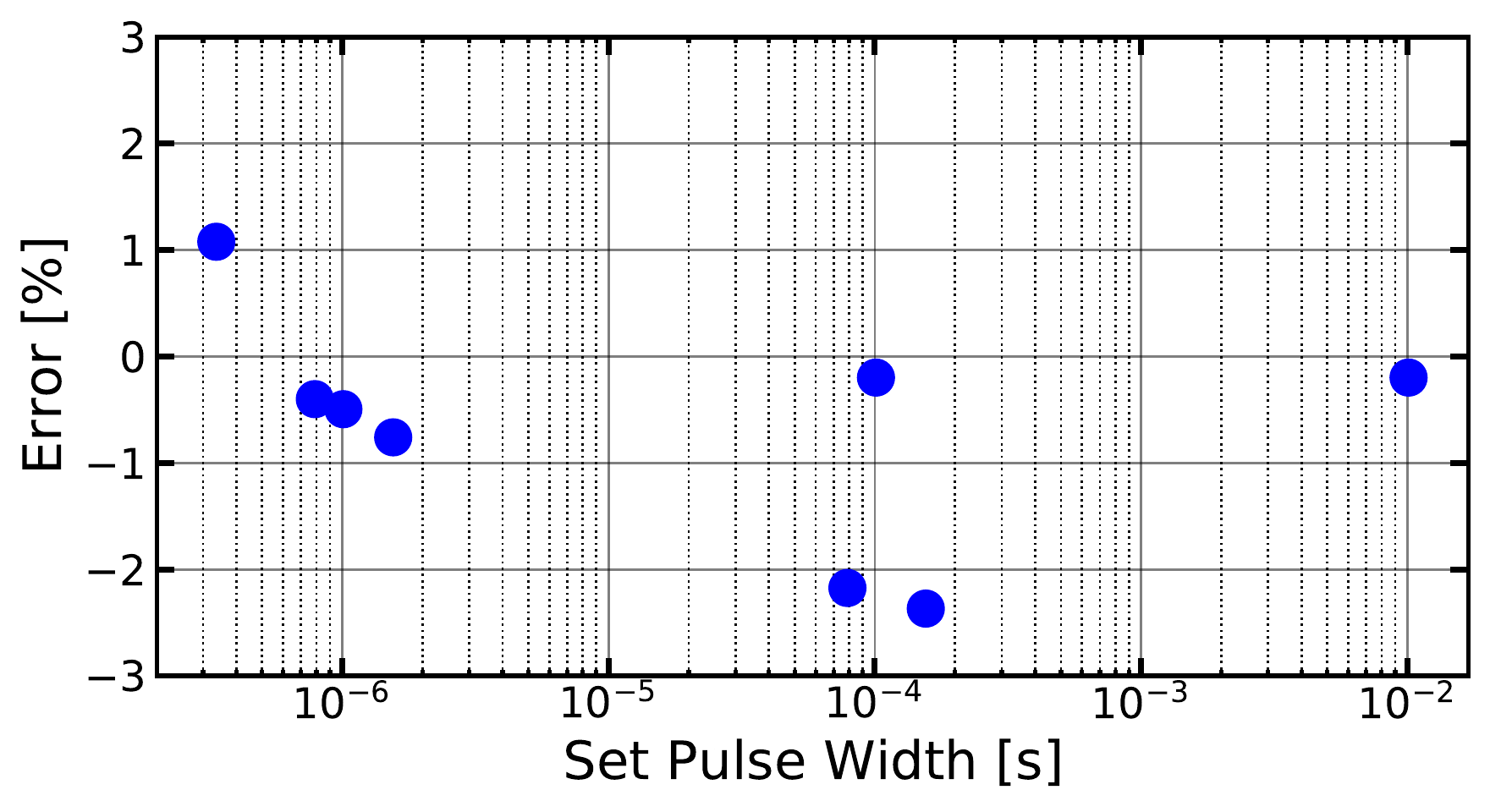}
    \subcaption{}
    \label{fig:measdataB}
  \end{subfigure}
  \caption{(\subref{fig:measdataA}) Difference between the measured and the simulated pulse time width using the nominal conditions. The dotted line indicates the ideal behavior. (\subref{fig:measdataB}) Error of the measured pulse time width after calibrating the value of the capacitor.}
\vspace{-0.5cm} 
\end{figure}

\begin{figure}[h]
  \begin{subfigure}{0.22\textwidth}
    \centering
    \includegraphics[width=0.95\textwidth]{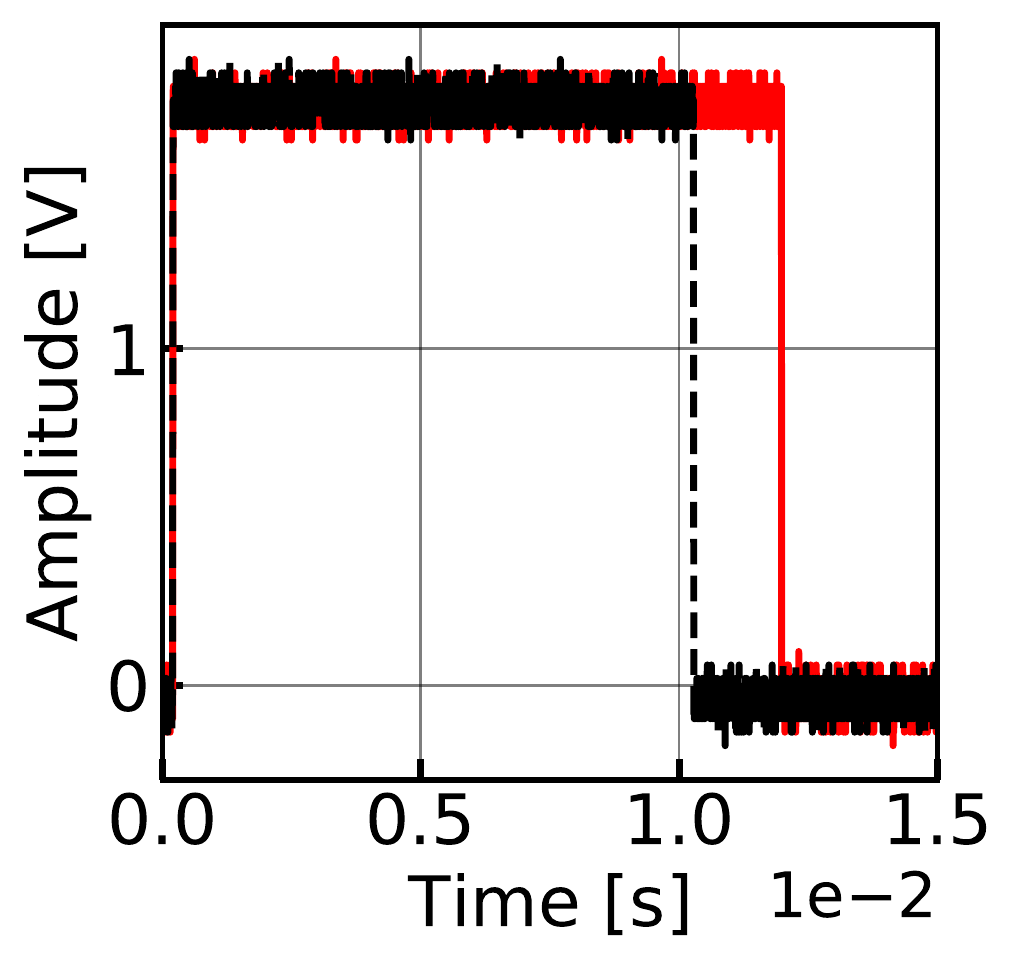}
    \subcaption{}
    \label{fig:10ms}
  \end{subfigure}
    \begin{subfigure}{0.22\textwidth}
    \centering
    \includegraphics[width=0.95\textwidth]{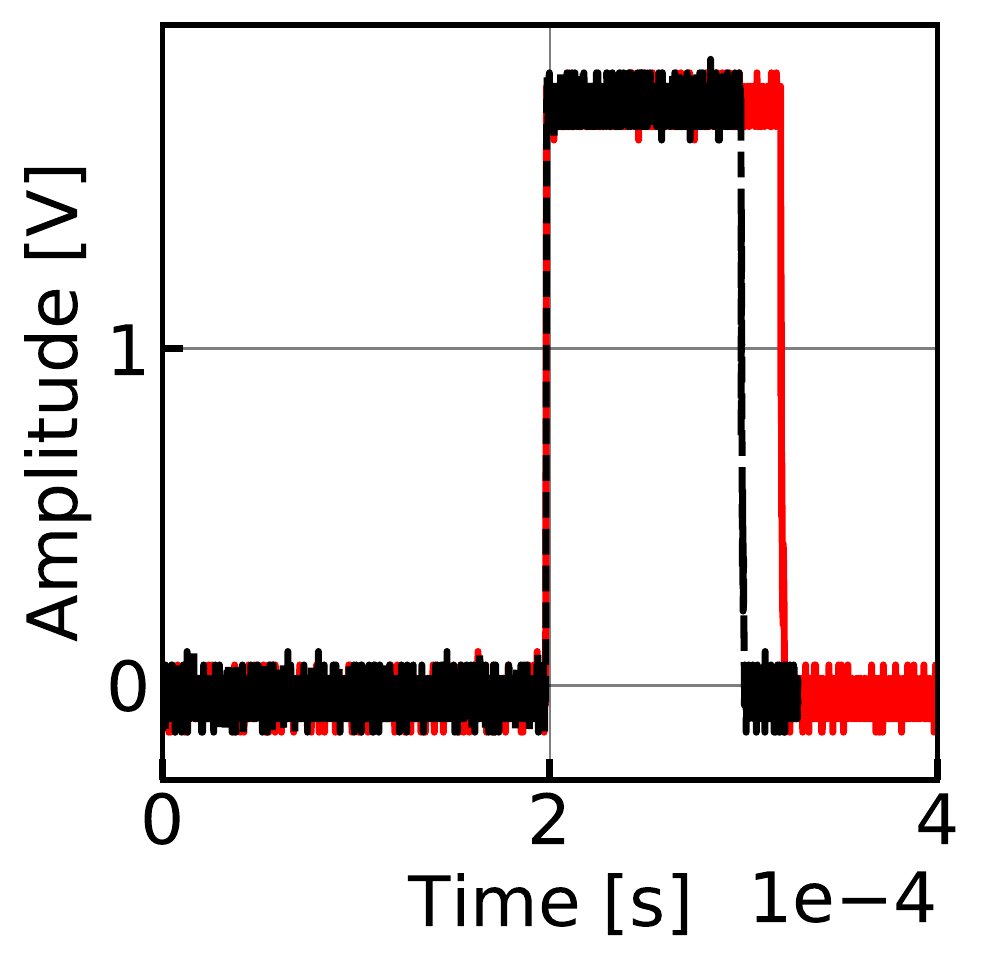}
    \subcaption{}
    \label{fig:100us}
  \end{subfigure}
  \\
  \begin{subfigure}{0.22\textwidth}
    \centering
    \includegraphics[width=0.95\textwidth]{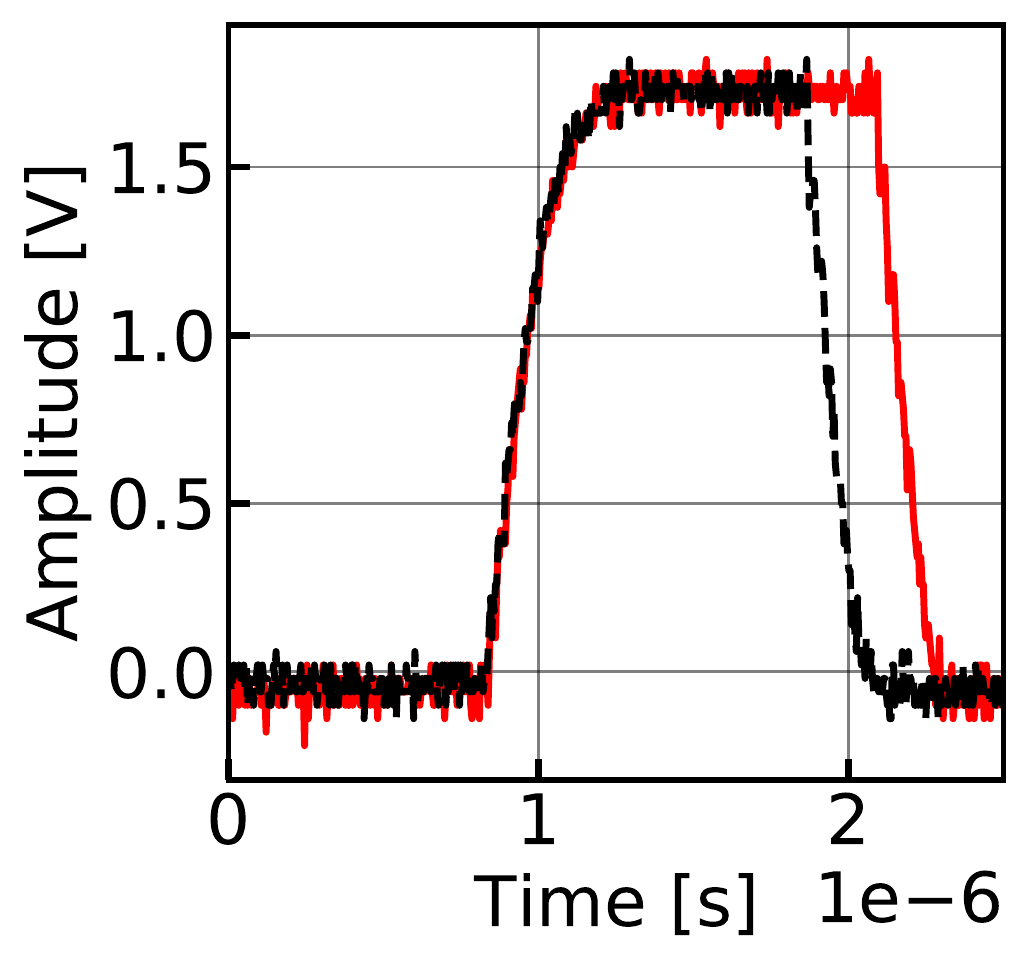}
    \subcaption{}
    \label{fig:1us}
  \end{subfigure}
  \begin{subfigure}{0.22\textwidth}
    \centering
    \includegraphics[width=0.95\textwidth]{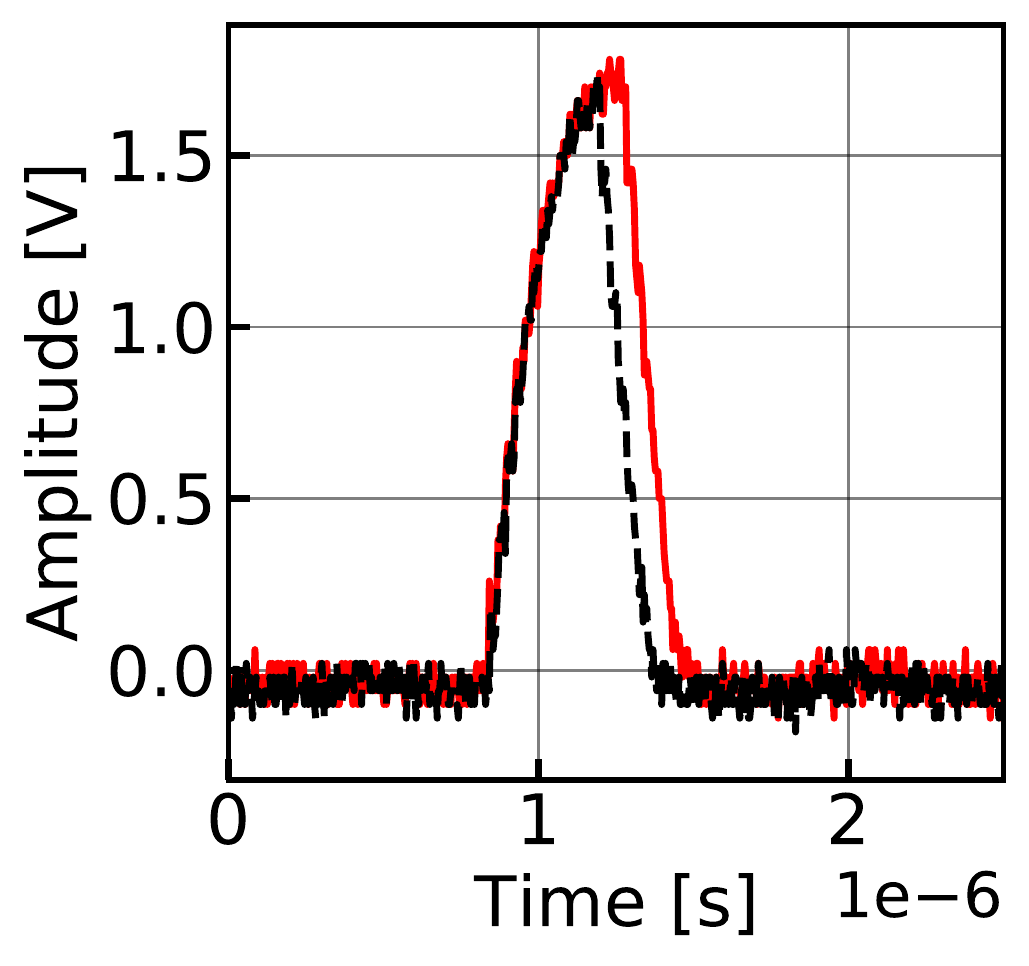}
    \subcaption{}
    \label{fig:300ns}
  \end{subfigure}
  % \vspace{10pt}
  \caption{Pre (red trace) and Post-calibrated (black dotted trace) pulse width measurements (\subref{fig:10ms}) 10\,ms; (\subref{fig:100us}) 100\,$\mu$s; (\subref{fig:1us}) 1\,$\mu$s (\subref{fig:300ns}) 336\,ns}
  \label{fig:measurments}  
\vspace{-0.5cm}  
\end{figure}

Initially, we biased the circuit using the same parameters used in the circuit simulations to produce pulses of desired widths. However, the fabricated circuits produced outputs that differ significantly from the circuit simulations (Fig.~\ref{fig:measdataA}). The option of precisely calibrating the capacitance value using the $C_{DAC}$ allows accurate pulse widths, with errors below $\pm2.5\%$ (Fig.~\ref{fig:measdataB}).
Here, only pulses down to 350\,ns width could be measured, due to the RC parasitic load induced by the setup. 
This is confirmed in Fig.~\ref{fig:measurments}, showing a selection of waveforms generated by the pulse generator under different conditions. 
The trend shown in Fig.~\ref{fig:measdataA} proves that the pulse generator can work at shorter pulse widths. In normal operating conditions, the device to be tested is placed right next to the Pulse Generator, whose photograph, as well as a detail of the contacts for the BEOL-integration of ferroelectric devices, are shown in Fig.~\ref{fig:pgphoto}. In this configuration, the integrated Pulse Generator is no longer limited by the parasitic load, and the generated waveform are expected to be adequate to precisely investigate the switching kinetics of ferroelectric devices.

\section{Conclusion}
We presented an on-chip pulse-generator designed and fabricated in 180\,nm technology. We demonstrated pulse generation across 6 orders of magnitude with an error of $\pm2.5\%$, while capable of driving relatively high parasitic capacitive loads present in the measurement setup. A low-parasitic on-chip interconnect between the the pulse-generator and the FeFET/FTJ devices placed on the same silicon ensures a reliable delivery of programming pulses with high accuracy. 
Future work includes exploring performance of the two DACs and their effect on the output pulses. This work is a key enabler for thorough characterization of the FeFETs/FTJs, otherwise limited to the parasitics of the measurement set up. 
\newpage
\bibliographystyle{IEEEtran}
\bibliography{Bibliography/biblioncs.bib, Bibliography/biblio_pg.bib}

\begin{thebibliography}{10}
\providecommand{\url}[1]{#1}
\csname url@samestyle\endcsname
\providecommand{\newblock}{\relax}
\providecommand{\bibinfo}[2]{#2}
\providecommand{\BIBentrySTDinterwordspacing}{\spaceskip=0pt\relax}
\providecommand{\BIBentryALTinterwordstretchfactor}{4}
\providecommand{\BIBentryALTinterwordspacing}{\spaceskip=\fontdimen2\font plus
\BIBentryALTinterwordstretchfactor\fontdimen3\font minus
  \fontdimen4\font\relax}
\providecommand{\BIBforeignlanguage}[2]{{%
\expandafter\ifx\csname l@#1\endcsname\relax
\typeout{** WARNING: IEEEtran.bst: No hyphenation pattern has been}%
\typeout{** loaded for the language `#1'. Using the pattern for}%
\typeout{** the default language instead.}%
\else
\language=\csname l@#1\endcsname
\fi
#2}}
\providecommand{\BIBdecl}{\relax}
\BIBdecl

\bibitem{Sony}
J.~Okuno, T.~Kunihiro, K.~Konishi, H.~Maemura, Y.~Shuto, F.~Sugaya,
  M.~Materano, T.~Ali, M.~Lederer, K.~Kuehnel, K.~Seidel, U.~Schroeder,
  T.~Mikolajick, M.~Tsukamoto, and T.~Umebayashi, ``High-endurance and
  low-voltage operation of 1t1c feram arrays for nonvolatile memory
  application,'' in \emph{2021 IEEE International Memory Workshop (IMW)}, 2021,
  pp. 1--3.

\bibitem{GF}
S.~Beyer, S.~Dünkel, M.~Trentzsch, J.~Müller, A.~Hellmich, D.~Utess, J.~Paul,
  D.~Kleimaier, J.~Pellerin, S.~Müller, J.~Ocker, A.~Benoist, H.~Zhou,
  M.~Mennenga, M.~Schuster, F.~Tassan, M.~Noack, A.~Pourkeramati, F.~Müller,
  M.~Lederer, T.~Ali, R.~Hoffmann, T.~Kämpfe, K.~Seidel, H.~Mulaosmanovic,
  E.~T. Breyer, T.~Mikolajick, and S.~Slesazeck, ``Fefet: A versatile cmos
  compatible device with game-changing potential,'' in \emph{2020 IEEE
  International Memory Workshop (IMW)}, 2020, pp. 1--4.

\bibitem{Benjamin2019}
B.~Max, M.~Hoffmann, S.~Slesazeck, and T.~Mikolajick, ``Direct correlation of
  ferroelectric properties and memory characteristics in ferroelectric tunnel
  junctions,'' \emph{IEEE Journal of the Electron Devices Society}, vol.~7, pp.
  1175--1181, 2019.

\bibitem{Erika-ISCAS}
E.~Covi, Q.~T. Duong, S.~Lancaster, V.~Havel, J.~Coignus, J.~Barbot,
  O.~Richter, P.~Klein, E.~Chicca, L.~Grenouillet, A.~Dimoulas, T.~Mikolajick,
  and S.~Slesazeck, ``Ferroelectric tunneling junctions for edge computing,''
  in \emph{2021 IEEE International Symposium on Circuits and Systems (ISCAS)},
  2021, pp. 1--5.

\bibitem{HalidSynapse}
H.~Mulaosmanovic, J.~Ocker, S.~Müller, M.~Noack, J.~Müller, P.~Polakowski,
  T.~Mikolajick, and S.~Slesazeck, ``Novel ferroelectric fet based synapse for
  neuromorphic systems,'' in \emph{2017 Symposium on VLSI Technology}, 2017,
  pp. T176--T177.

\bibitem{Mattia}
\BIBentryALTinterwordspacing
M.~Halter, L.~Bégon-Lours, V.~Bragaglia, M.~Sousa, B.~J. Offrein, S.~Abel,
  M.~Luisier, and J.~Fompeyrine, ``Back-end, cmos-compatible ferroelectric
  field-effect transistor for synaptic weights,'' \emph{ACS Applied Materials
  \& Interfaces}, vol.~12, no.~15, pp. 17\,725--17\,732, 2020, pMID: 32192333.
  [Online]. Available: \url{https://doi.org/10.1021/acsami.0c00877}
\BIBentrySTDinterwordspacing

\bibitem{HalidNeuron}
B.~Suresh, M.~Bertele, E.~T. Breyer, P.~Klein, H.~Mulaosmanovic, T.~Mikolajick,
  S.~Slesazeck, and E.~Chicca, ``Simulation of integrate-and-fire neuron
  circuits using hfo<inf>2</inf>-based ferroelectric field effect
  transistors,'' in \emph{2019 26th IEEE International Conference on
  Electronics, Circuits and Systems (ICECS)}, 2019, pp. 229--232.

\bibitem{Indiveri_Liu15}
G.~Indiveri and S.-C. Liu, ``Memory and information processing in neuromorphic
  systems,'' \emph{Proceedings of the {IEEE}}, vol. 103, no.~8, pp. 1379--1397,
  2015.

\bibitem{kinetics}
\BIBentryALTinterwordspacing
H.~Mulaosmanovic, J.~Ocker, S.~Müller, U.~Schroeder, J.~Müller,
  P.~Polakowski, S.~Flachowsky, R.~van Bentum, T.~Mikolajick, and S.~Slesazeck,
  ``Switching kinetics in nanoscale hafnium oxide based ferroelectric
  field-effect transistors,'' \emph{ACS Applied Materials \& Interfaces},
  vol.~9, no.~4, pp. 3792--3798, 2017, pMID: 28071052. [Online]. Available:
  \url{https://doi.org/10.1021/acsami.6b13866}
\BIBentrySTDinterwordspacing

\bibitem{materano2020polarization}
M.~Materano, P.~D. Lomenzo, H.~Mulaosmanovic, M.~Hoffmann, A.~Toriumi,
  T.~Mikolajick, and U.~Schroeder, ``Polarization switching in thin doped hfo2
  ferroelectric layers,'' \emph{Applied Physics Letters}, vol. 117, no.~26, p.
  262904, 2020.

\bibitem{sw-ret}
H.~Mulaosmanovic, F.~Müller, M.~Lederer, T.~Ali, R.~Hoffmann, K.~Seidel,
  H.~Zhou, J.~Ocker, S.~Mueller, S.~Dünkel, D.~Kleimaier, J.~Müller,
  M.~Trentzsch, S.~Beyer, E.~T. Breyer, T.~Mikolajick, and S.~Slesazeck,
  ``Interplay between switching and retention in hfo<sub>2</sub>-based
  ferroelectric fets,'' \emph{IEEE Transactions on Electron Devices}, vol.~67,
  no.~8, pp. 3466--3471, 2020.

\bibitem{covi2014TSM}
E.~Covi, A.~Cabrini, L.~Vendrame, L.~Bortesi, R.~Gastaldi, and G.~Torelli,
  ``On-wafer analog pulse generator for fast characterization and parametric
  test of resistive switching memories,'' \emph{IEEE Transactions on
  Semiconductor Manufacturing}, vol.~27, no.~2, pp. 134--150, 2014.

\bibitem{weste2011cmos}
N.~Weste, \emph{CMOS VLSI Design: A Circuits and Systems Perspective}.\hskip
  1em plus 0.5em minus 0.4em\relax Addison Wesley, 2011.

\bibitem{AllstotComparator}
D.~Allstot, ``A precision variable-supply cmos comparator,'' \emph{IEEE Journal
  of Solid-State Circuits}, vol.~17, no.~6, pp. 1080--1087, 1982.

\bibitem{allen2011cmos}
P.~Allen and D.~Holberg, \emph{CMOS Analog Circuit Design}, ser. The Oxford
  Series in Electrical and Computer Engineering.\hskip 1em plus 0.5em minus
  0.4em\relax OUP USA, 2011.

\bibitem{LeePG}
S.~Lee and H.~Yoo, ``Configurable digital pulse generator for neuromorphic
  devices,'' in \emph{2021 International Conference on Electronics,
  Information, and Communication (ICEIC)}, 2021, pp. 1--3.

\bibitem{ErfanijaziPG}
H.~Erfanijazi, T.~Serrano-Gotarredona, and B.~Linares-Barranco, ``Novel
  programmable single pulse generator for producing pulse widths in different
  time scales,'' in \emph{2021 International Conference on Content-Based
  Multimedia Indexing (CBMI)}, 2021, pp. 1--6.

\end{thebibliography}
\end{document}